\documentclass[sigconf]{acmart}

\copyrightyear{2022}
\acmYear{2022}
\setcopyright{acmcopyright}\acmConference[MM '22]{Proceedings of the 30th ACM International Conference on Multimedia}{October 10--14, 2022}{Lisboa, Portugal}
\acmBooktitle{Proceedings of the 30th ACM International Conference on Multimedia (MM '22), October 10--14, 2022, Lisboa, Portugal}
\acmPrice{15.00}
\acmDOI{10.1145/3503161.3548025}
\acmISBN{978-1-4503-9203-7/22/10}

\acmSubmissionID{mmfp1163}

\usepackage{multirow}
\usepackage{subfigure}
\usepackage{float}
\usepackage{balance}

\begin{document}

\title{CubeMLP: An MLP-based Model for Multimodal Sentiment Analysis and Depression Estimation}

\author{Hao Sun}
\email{sunhaoxx@zju.edu.cn}
\orcid{0000-0001-8094-1991}
\affiliation{
  \department{College of Computer Science and Technology}
  \institution{Zhejiang University}
  \city{Hangzhou}
  \country{China}
}

\author{Hongyi Wang}
\email{whongyi@zju.edu.cn}
\orcid{0000-0002-2336-1496}
\affiliation{
  \department{College of Computer Science and Technology}
  \institution{Zhejiang University}
  \city{Hangzhou}
  \country{China}
}

\author{Jiaqing Liu}
\email{liu-j@fc.ritsumei.ac.jp}
\orcid{0000-0002-5340-7053}
\affiliation{
  \department{College of Information Science and Engineering}
  \institution{Ritsumeikan University}
  \city{Shiga}
  \country{Japan}
}

\author{Yen-Wei Chen}
\email{chen@is.ritsumei.ac.jp}
\orcid{0000-0002-5952-0188}
\affiliation{
  \department{College of Information Science and Engineering}
  \institution{Ritsumeikan University}
  \city{Shiga}
  \country{Japan}
}

\author{Lanfen Lin}
\email{llf@zju.edu.cn}
\orcid{0000-0003-4098-588X}
\affiliation{
  \department{College of Computer Science and Technology}
  \institution{Zhejiang University}
  \city{Hangzhou}
  \country{China}
}

\renewcommand{\shortauthors}{Hao Sun et al.}

\begin{abstract}
Multimodal sentiment analysis and depression estimation are two important research topics that aim to predict human mental states using multimodal data.
Previous research has focused on developing effective fusion strategies for exchanging and integrating mind-related information from different modalities.
Some MLP-based techniques have recently achieved considerable success in a variety of computer vision tasks.
Inspired by this, we explore multimodal approaches with a \textit{feature-mixing} perspective in this study. 
To this end, we introduce CubeMLP, a multimodal feature processing framework based entirely on MLP.
CubeMLP consists of three independent MLP units, each of which has two affine transformations.
CubeMLP accepts all relevant modality features as input and \textit{mixes} them across three axes.
After extracting the characteristics using CubeMLP, the mixed multimodal features are flattened for task predictions.
Our experiments are conducted on sentiment analysis datasets: CMU-MOSI and CMU-MOSEI, and depression estimation dataset: AVEC2019.
The results show that CubeMLP can achieve state-of-the-art performance with a much lower computing cost.
\end{abstract}



\maketitle

\section{Introduction}
Multimodal data has become an important means of communication for individuals and the public as social media has grown in prevalence.
In this scenario, estimating human psyche states from multimodal data, such as sentiment tendencies and depression levels, becomes increasingly important.
Multimodal data commonly includes textual ($t$), acoustic($a$), and visual ($v$) information.
The characteristics of features can be extracted from the multimodal data as two-dimensional matrices $R^{L_{m}\times D_{m}}$, where $L_{m}$ and $D_{m}$ are the sequential length and feature-channel size of modality $m$, respectively.

To effectively process multimodal features, Zadeh et al.~\cite{zadeh2017tensor} first introduced Cartesian product in Tensor Fusion Network (TFN) to blend the features from all involved modalities at $L_{m}$ axis.
Many researchers then proposed that there are bidirectional relationships and complementary information among modalities, and apply the attention mechanism to calculate the coattention in modality pairs (like textual and acoustic)~\cite{zadeh2018multi,gu2018multimodal}.
Most recently, with the remarkable success of Transformer-based structures~\cite{vaswani2017attention}, some works attempt to employ the self-attention mechanism for modality interactions~\cite{delbrouck2020transformer,wang2020transmodality,delbrouck2020modulated,han2021bi}.
The core of these trending methods is mostly the exchange of information between modalities.

The consequences of these information-exchanging methods could be viewed as feature \textit{mixing}.
For instance, the TFN~\cite{zadeh2017tensor} and some sequential-wise coattention approaches~\cite{zadeh2018multi,li2020multimodal} attempted to mix the features on $L_{m}$ axis between modalities, while the channel-wise coattention approaches~\cite{mai2021analyzing} tried to perform the mixing on $D_{m}$ axis.
As for the Transformer-based approaches~\cite{delbrouck2020transformer,wang2020transmodality,delbrouck2020modulated,han2021bi}, they employ the self-attention mechanism and perform the sophisticated mixing on $L_{m}$ axis between paired modalities, which can also be treated as enhancing one modality by the other.

A number of variations, such as ViT~\cite{dosovitskiy2020image} and ViViT~\cite{arnab2021vivit} have been proposed recently as a result of the increasing use of transformers in computer vision applications. 
Transformers, on the other hand, have a significant memory requirement for self-attention, which is a major drawback of the Transformer's architecture.
Consequently, structures composed entirely of multilayer perceptrons (MLPs) have gained interest. 
MLP-mixer~\cite{tolstikhin2021mlp} and ResMLP~\cite{touvron2021resmlp}, for example, use MLPs to replace the self-attention mechanism design. 
By substituting self-attention with MLPs, these techniques significantly reduce computational costs while maintaining high performance.

Inspired by MLP-based techniques, here, we propose CubeMLP, a simple yet effective MLP-based framework for multimodal feature processing.
During preprocessing, we integrate the modality features into a multimodal tensor $X \in R^{L\times M \times D}$ where $M$ is the number of modalities, $L$ is sequential length and $D$ is the size of feature channels.
CubeMLP is composed of three MLP units for three respective axis ($L$, $M$, and $D$).
The first MLP unit is designed to mix up features on the $L$ axis, a process called \textit{sequential-mixing}.
\textit{Modality-mixing} ($t$, $a$, and $v$) is performed by the second MLP unit on the $M$ axis.
Finally, the third MLP unit on the $D$ axis performs \textit{channel-mixing}.
Every MLP unit contains two fully-connected layers, each of which contains an affine transformation that can be represented mathematically as a matrix $W$ with bias $B$.
In CubeMLP, we mix multimodal features on each possible axis using the proposed three MLP design structures.
After that, the mixed features are flattened and fed to the classifier for the predictions.
During this procedure, the multimodal features are fused and multimodal information is exchanged on any axis.

Our contributions can be summarized as follows:
\begin{itemize}
\item We propose CubeMLP, a multimodal feature processing framework based exclusively on MLP.
CubeMLP mixes features on three axes: sequence ($L$), modality ($M$), and channel ($D$).
The distinct multimodal information ($t$, $a$, and $v$) is effectively transmitted and shared during the mixing process to extract important features for sentiment analysis and depression detection.
\item We propose to use MLPs to largely reduce the computational burden while achieving competitive results with some state-of-the-art approaches, which proves that our CubeMLP is an efficient structure for multimodal feature processing.
\item We conduct thorough experiments on two mind state estimation tasks to validate the effectiveness of CubeMLP: multimodal sentiment analysis and multimodal depression detection.
The results show that our method is favorably competitive with the state-of-the-art methods for sentiment analysis, while achieving great progress for depression detection.
\end{itemize}

\section{Related Works}
\subsection{Multimodal Sentiment Analysis}
The task of multimodal sentiment analysis is to predict the sentiment tendencies from human's facial expression($v$), acoustic tone($a$), and spoken words($t$) in each utterance.
Zadeh et al. firstly proposed Tensor Fusion Network (TFN)~\cite{zadeh2017tensor} and Memory Fusion Network (MFN)~\cite{zadeh2018memory} to fuse the multimodal features on sequential level.
Recent works are mostly committed to calculate correlations among involved modalities.
For example, Chen et al.~\cite{chen2020swafn} developed a Sentimental Words Aware Fusion Network (SWAFN) to compute the coattention between text and other modalities.
Deng et al.~\cite{deng2021dense} proposed a deep dense fusion network with multimodal residual (DFMR) to integrate multimodal information in a paired manner.
With the significant progress of Transformer~\cite{vaswani2017attention} in natural language processing and computer vision, some works~\cite{wang2020transmodality,delbrouck2020modulated,han2021bi} applied its self-attention mechanism to perform modality interactions.
For example, Delbrouck~\cite{delbrouck2020transformer} designed a Transformer-based joint-encoding (TBJE) that takes acoustic and textual features as input and jointly encode two modalities.
Tsai~\cite{tsai2019multimodal} proposed a Multimodal Transformer (MulT) and attempted to calculate cross-modal attention between paired modalities($t$ with $a$, for instance).

\subsection{Multimodal Depression Detection}
Compared to multimodal sentiment analysis, depression detection requires longer time sequences because it is a persistent long-term characteristic obtained from a human.
Joshi et al.~\cite{joshi2013multimodal} used the bag-of-words model to encode acoustic and visual features and then fused them via principal component analysis (PCA) and support vector mechanisms (SVM). 
Rodrigues et al.~\cite{rodrigues2019multimodal} used audio-translated texts with their hidden embedding extracted from pretrained BERT~\cite{devlin2018bert} model, and employ CNNs to obtain cross-modality information. 
Kaya et al.~\cite{kaya2019predicting} designed a new Automatic Speech Recognizer (ASR) transcription based features while Ray~\cite{ray2019multi} proposed a multi-layer attention network for estimating depressions.
Aside from acoustic, visual, and textual features, Kroenke et al.~\cite{kroenke2002phq} showed that body gestures have a positive contribution to the accuracy of depression estimation.
Furthermore, Sun et al.~\cite{sun2021multi} employed Transformer model to extract multimodal features and proposed an adaptive late fusion scheme for the final predictions.
Zhao~\cite{zhao2020hybrid} introduced a hybrid feature extraction structure which combines self-attention and 3D convolutions for different kinds of features.

\subsection{MLP-based Models}
MLP-based models are new proposed structures for vision tasks.
In ViT~\cite{dosovitskiy2020image}, Transformer is firstly used for image processing.
It splits images into several patches and feeds them to Transformer.
The significant performance boost prompts many other variations~\cite{arnab2021vivit,liu2021swin}, but the computational burden is still huge.
Therefore, some MLP-based models are proposed including MLP-Mixer~\cite{tolstikhin2021mlp}, ResMLP~\cite{touvron2021resmlp}, and Hire-MLP~\cite{guo2021hire}.
These methods discard the self-attention mechanism and replace with MLPs, which is more efficient.
In general, two independent MLPs are included in these models, one MLP processes channels while the other processes the tokens.

Inspired of the inherent structure of MLP-based models, we find it can be naturally transferred to multimodal feature processing.
Because multimodal features usually consist of three axis (sequential, modality, and channel), we add an additional MLP to comprehensively mix the features.
The detailed structures are illustrated in Section~\ref{method}.

\begin{figure*}[h]
  \centering
  \includegraphics[width=\linewidth]{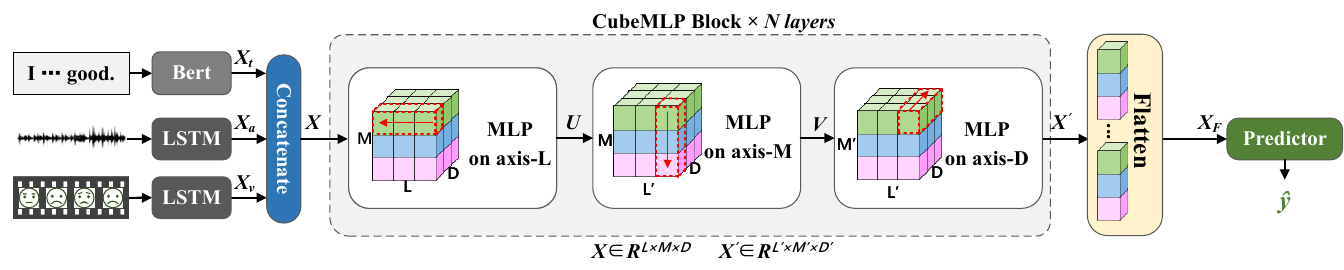}
  \caption{The overview of the MLP-based mixer. The extracted features are fed to $N$ layer stacked CubeMLP blocks to be mixed. The mixed features are flattened for the prediction.}
  \label{fig_overview}
  \Description{}
\end{figure*}

\begin{figure}[h]
  \centering
  \includegraphics[width=1.0\linewidth]{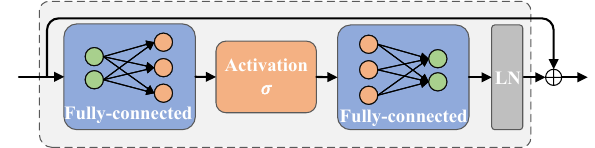}
  \caption{The structure of MLP unit. \textit{LN} means the layer normalization. A residual short-cut is employed in the unit.}
  \label{fig_mlp}
  \Description{}
\end{figure}
\section{Proposed Methods}~\label{method}
CubeMLP is a simple but effective multimodal feature processing structure.
Our task is to predict sentiment tendency or depression level from human utterances in videos, where each utterance is an input sample to the model.
In one utterance, three modalities are provided including textual($t$), acoustic($a$) and visual($v$).
The overview of our approach is shown in Figure~\ref{fig_overview}.
Each modality features are first extracted by specific methods.
After features being extracted, we don't exchange cross-modality information on sequential level or channel level, as in previous approaches~\cite{delbrouck2020modulated,zadeh2018multimodal}.
Instead, we perform the mixing individually on sequential, channel, and modality levels.
Specifically, the CubeMLP is used to \textit{mix up} multimodal features on all axis by respective MLP units.
The mixed multimodal features are then passed to the classifier to perform the predictions for sentiment analysis or depression detection.

\subsection{Feature Extraction}
Before mixing the modalities, LSTM~\cite{hochreiter1997long} is employed to extract acoustic and visual features while pretrained BERT~\cite{devlin2018bert} is used for extracting textual features.
The extracted features are represented as $X_{m} \in R^{L_{}\times D_{}} $ in which $L_{}$ denotes the length of the utterance, $D_{}$ denotes the respective feature channel dimension and  $m \in \{t, a, v\}$.
The extracted modality features share the same sequential lengths and feature channel dimensions.
The goal is to mix the modality features $X_{t}$, $X_{a}$, and $X_{v}$, and predict the continuous affective tendency $\hat{y} \in R$ for each utterance.

\subsection{The Structure of CubeMLP}
After the features being extracted, we first extend $X_{m}\in R^{L \times D}$ at the second dimension into $R^{L \times 1 \times D}$ and then concatenate them to comprise the multimodal features $X \in R^{L\times M \times D}$ along the extended axis, where $M$ is the number of modalities.
The multimodal features are then passed to $N$ layer stacked CubeMLPs to be mixed-up as shown in Figure~\ref{fig_overview}.
The CubeMLP block consists of three MLP units, and each MLP unit is designed to \textit{mix} the multimodal features on its respective axis.
Specifically, the first MLP $f_{L}:R^{L\times *\times *}\rightarrow R^{L'\times *\times *}$ aims to perform \textit{sequential-mixing} which acts on $L$ axis.
The second one  $f_{M}:R^{*\times M\times *}\rightarrow R^{*\times M'\times *}$ is \textit{modality-mixing} MLP and acts on $M$ axis.
The third MLP $f_{D}:R^{*\times*\times D}\rightarrow R^{*\times *\times D'}$ executes the \textit{channel-mixing} on $D$ axis.
Each MLP unit is composed of two fully-connected layers and a nonlinear activation as shown in Figure~\ref{fig_mlp}.
The fully-connected layers can also be treated as two affine transformations.

Let consider the first sequential-mixing MLP unit on $L$ axis.
Tensor $X$ can be seen as a set of vectors of $X_{*,m,d}\in R^{L\times 1\times 1}$, where $(m,d) \in \{(1,1), (1,2),...,(2,1), (2,2),..., (M,D) \}$.
$X_{*,m,d}$ is the vector of $m$-th modality and $d$-th channel.
Each affine transformation in the sequential-mixing MLP can be represented as:
\begin{equation}
  \begin{aligned}
    Aff_{L}(X_{*,m,d}) = W_{L}X_{*,m,d} + B_{L}, \\
  \end{aligned}
  \label{equ:affine}
\end{equation}
where $W_{L} \in R^{L\times L'}$ and $B_{L} \in R^{L'}$ are two matrix-represented learnable parameters.
$L'$ is the reduced dimension on $L$-axis, which is a hyperparameter and will be discussed in more detail in Section~\ref{dim_explore}.
Equation~\ref{equ:affine} indicates that all $X_{*,m,d}$ share parameters $W_{L}$ and $B_{L}$.
Therefore, the complete MLP unit can be mathematically represented as:
\begin{equation}
  \begin{aligned}
    U_{*,m,d} = LN(Aff_{L}(\sigma(Aff_{L}(X_{*,m,d})))+X_{*,m,d}) \in R^{L'\times 1\times 1},&\\
    {\rm for}\ (m,d) \in \{(1,1), (1,2),...,(2,1), (2,2),..., (M,D)\},&  \\
  \end{aligned}
  \label{equ:mlp}
\end{equation}
where $\sigma$ is the nonlinear activation, $LN$ is the layer normalization.
The output tensor $U\in R^{L'\times M\times D}$ of the first MLP unit can be considered as a set of vectors of $U_{*,m,d} \in R^{L'\times 1\times 1}$, where $(m,d) \in \{(1,1), (1,2),...,(2,1), (2,2),..., (M,D)\}$.

As well as the first MLP unit on the axis of $L$, the output $V\in R^{L' \times M' \times D}$ of the second MLP on $M$-axis and the output $X'\in R^{L' \times M' \times D'}$ of the third MLP on $D$-axis can be considered as a set of their elementary vectors $V_{l,*,d} \in R^{1\times M'\times 1}$ and $X'_{l,m,*} \in R^{1\times 1\times D'}$, where $M'$ and $D'$ are reduced dimensions on $M$-axis and $D$-axis, respectively, which will 
also be discussed in Section~\ref{dim_explore}. 
As well as Equation~\ref{equ:mlp}, $V_{l,*,d}$ and $X'_{l,m,*}$ can be calculated as Equation~\ref{equ:mlpm} and~\ref{equ:mlpd}, respectively.
\begin{equation}
  \begin{aligned}
    V_{l,*,d} = LN(Aff_{M}(\sigma(Aff_{M}(U_{l,*,d})))+U_{l,*,d}) \in R^{1\times M'\times 1},&\\
    {\rm for}\ (l,d) \in \{(1,1), (1,2),...,(2,1), (2,2),..., (L',D) \},& \\
  \end{aligned}
  \label{equ:mlpm}
\end{equation}
\begin{equation}
  \begin{aligned}
    X^{'}_{l,m,*} = LN(Aff_{D}(\sigma(Aff_{D}(V_{l,m,*})))+V_{l,m,*}) \in R^{1\times 1\times D'},&\\
    {\rm for}\ (l,m) \in \{(1,1), (1,2),...,(2,1), (2,2),..., (L',M') \}.&
  \end{aligned}
  \label{equ:mlpd}
\end{equation}
The $X^{'} \in R^{L'\times M'\times D'}$ is the mixed multimodal features.

\subsection{Prediction Head}
Following the previous works~\cite{han2021bi,zadeh2017tensor}, the mixed multimodal features $X^{'} \in R^{L'\times M'\times D'}$ are flattened as $X^{'} \in R^{L'M'D'}$.
Then the flattened features are fed to the classifier $f_{c}:R^{L'M'D'}\rightarrow R$ to predict the sentiment tendency or depression level.

We use the mean absolute error (MAE) as the loss function during training for multimodal sentiment analysis, which is a regression task:
\begin{equation}
  \begin{aligned}
    \mathcal{L}_{mae} = \frac{1}{N}\sum_{i=1}^{N}|y_{i} - \hat{y}_{i}|  , \\
  \end{aligned}
  \label{equ:mae}
\end{equation}
where $N$ is the number of samples.
The absolute errors between the prediction and the ground truth are calculated using MAE. 
MAE has a substantial effect on small errors than higher-order errors, allowing models to achieve better accuracy on subtle sentiments.
As a result, MAE is often used as the key performance metric for sentiment analysis~\cite{han2021bi,zadeh2017tensor,chen2020swafn,hazarika2020misa}.

For depression detection, we train the model to regress the depression tendency with Concordance Correlation Coefficient (CCC) loss as the loss cost:
\begin{equation}
  \begin{aligned}
    \mathcal{L}_{CCC} = 1.0 - \frac{2S_{\hat{y}y}}{S_{\hat{y}}^{2}+S_{y}^{2} + (\bar{\hat{y}} - \bar{y})^{2}} 
  \end{aligned}
  \label{equ:cccloss}
\end{equation}
CCC is widely used for estimating depressions~\cite{joshi2013multimodal,rodrigues2019multimodal}, because it is not only unbiased by changes in scale and location, but also includes measure on both correlation and accuracy~\cite{lawrence1989concordance}.

\begin{table*}[]
\caption{The results on two multimodal sentiment analysis benchmark datasets, CMU-MOSI and CMU-MOSEI.}
\label{result_msa}
\centering
    \begin{tabular}{c|ccccc|ccccc}
    \toprule
    \multirow{2}{*}{Models} & \multicolumn{5}{c|}{CMU-MOSI}     & \multicolumn{5}{c}{CMU-MOSEI}\\
    & MAE($\downarrow$) & Corr($\uparrow$) & Acc-2($\uparrow$) & F1-Score($\uparrow$) & Acc-7($\uparrow$) & MAE($\downarrow$) & Corr($\uparrow$) & Acc-2($\uparrow$) & F1-Score($\uparrow$) & Acc-7($\uparrow$) \\
    \midrule
    TFN\cite{zadeh2017tensor}         & 0.970 & 0.633 & 73.9 & 73.4 & 32.1 & 0.593 & 0.700 & 82.5 & 82.1 & 50.2 \\
    MFN\cite{zadeh2018memory}         & 0.965 & 0.632 & 77.4 & 77.3 & 34.1 &  -    & -     & 76.0 & 76.0 & -    \\
    ICCN\cite{sun2020learning}        & 0.862 & 0.714 & 83.0 & 83.0 & 39.0 & 0.565 & 0.713 & 84.2 & 84.2 & 51.6 \\
    SWAFN\cite{chen2020swafn}         & 0.880 & 0.697 & 80.2 & 80.1 & 40.1 & -     & -     & -    & -    & -   \\
    MulT\cite{tsai2019multimodal}     & 0.871 & 0.698 & 83.0 & 82.8 & 40.0 & 0.580 & 0.703 & 82.5 & 82.3 & 51.8 \\
    LMF-MulT\cite{sahay2020low}       & 0.957 & 0.681 & 78.5 & 78.5 & 34.0 & 0.620 & 0.668 & 80.8 & 81.3 & 49.3 \\
    MAT\cite{delbrouck2020transformer}& -     & -     & -    & 80.0 & -    & -     & -     & 82.0 & 82.0 & -    \\
    MNT\cite{delbrouck2020transformer}& -     & -     & -    & 80.0 & -    & -     & -     & 80.5 & 80.5 & -    \\
    MISA\cite{hazarika2020misa}       & 0.817 & 0.748 & 82.1 & 82.0 & 41.4 & 0.557 & 0.748 & 84.9 & 84.8 & 51.7 \\
    BBFN\cite{han2021bi}              & 0.776 & 0.755 & 84.3 & 84.3 & 45.0 & \textbf{0.529} & \textbf{0.767} & \textbf{86.2} & \textbf{86.1} & 54.8 \\
    CubeMLP(Ours)                     & \textbf{0.770} & \textbf{0.767} & \textbf{85.6} & \textbf{85.5} & \textbf{45.5} & \textbf{0.529} & 0.760 & 85.1 & 84.5 & \textbf{54.9}\\
    \bottomrule
    \end{tabular}
\end{table*}

\begin{table}[]
\caption{The results on the test set of depression detection benchmark dataset, AVEC2019 DDS.}
\label{result_dds}
\centering
    \begin{tabular}{c|cc}
    \toprule
    \multirow{2}{*}{Models} & \multicolumn{2}{c}{AVEC2019}  \\
    & CCC($\uparrow$) & MAE($\downarrow$)  \\
    \midrule
    Baseline~\cite{ringeval2019avec}                      & 0.111 & 6.37 \\
    Adaptive Fusion Transformer~\cite{sun2021multi}       & 0.331 & 6.22 \\
    EF~\cite{kaya2019predicting}                          & 0.344 & -    \\
    Bert-CNN \& Gated-CNN~\cite{rodrigues2019multimodal}  & 0.403 & 6.11 \\
    Multi-scale Temporal Dilated CNN~\cite{fan2019multi}  & 0.430 & 4.39 \\
    Hierarchical BiLSTM~\cite{yin2019multi}               & 0.442 & 5.50 \\
    CubeMLP(Ours)                       & \textbf{0.583} & \textbf{4.37} \\
    \bottomrule
    \end{tabular}
\end{table}

\begin{table*}[htbp]
  \centering
  \caption{The ablation study of our methods on CMU-MOSI~\cite{zadeh2016multimodal}. MLP-L means the sequential-mixing, MLP-M means the modality-mixing, and MLP-D means the channel mixing.}
  \label{tab:result_ablation}
  \begin{tabular}{cccc|ccccc}
    \toprule
    & MLP-L & MLP-M & \multicolumn{1}{c|}{MLP-D} & MAE($\downarrow$) & Corr($\uparrow$) & Acc-2($\uparrow$) & F1-Score($\uparrow$) & Acc-7($\uparrow$) \\
    \hline
    Model 1 & \checkmark &           &            & 0.860 & 0.744 & 80.6 & 80.7 & 39.0 \\
    Model 2 &            &\checkmark &            & 0.850 & 0.729 & 80.3 & 80.4 & 39.2  \\
    Model 3 &            &           &\checkmark  & 0.910 & 0.717 & 81.7 & 81.8 & 39.0  \\
    Model 4 & \checkmark &\checkmark &            & 0.806 & 0.753 & 81.5 & 81.6 & 42.4  \\
    Model 5 & \checkmark &           &\checkmark  & 0.803 & 0.750 & 80.6 & 80.8 & 39.5  \\
    Model 6 &            &\checkmark & \checkmark & 0.874 & 0.718 & 82.4 & 82.4 & 41.6  \\
    Model 7(Ours) & \checkmark &\checkmark &\checkmark &\textbf{0.770} &\textbf{ 0.767} &\textbf{85.6} &\textbf{85.5} &\textbf{45.5}  \\
    \bottomrule
  \end{tabular}
\end{table*}

\section{Experiments}
\subsection{Datasets}
We conduct the experiments on two multimodal mind state estimation tasks: sentiment analysis and depression detection, as previous works find that there is a strong connection between them~\cite{qureshi2020improving}.
For multimodal sentiment analysis, we use two popular benchmark datasets: CMU-MOSI~\cite{zadeh2016multimodal} and CMU-MOSEI~\cite{zadeh2018multimodal}.
For multimodal depression detection, we use the AVEC2019 dataset~\cite{ringeval2019avec} to evaluate the effectiveness of CubeMLP.
\subsubsection{CMU-MOSI}
CMU-MOSI~\cite{zadeh2016multimodal} is a popular dataset for multimodal sentiment analysis.
The samples in CMU-MOSI are utterance-based videos that are collected from the Internet.
In each sample the speakers express their subjective opinions on certain topics.
1283 training utterances, 229 validation utterances, and 686 test utterances are provided in the dataset.
The dataset is annotated with sentiment tendencies in the interval $[-3, 3]$.
\subsubsection{CMU-MOSEI}
CMU-MOSEI~\cite{zadeh2018multimodal} is enlarged from the CMU-MOSI.
It has the same annotations as the CMU-MOSI.
In CMU-MOSEI, there are 16315 utterances for training, 1817 utterances for validation, and 4654 utterances for testing.
\subsubsection{AVEC2019}
The AVEC2019 DDS dataset~\cite{ringeval2019avec} is obtained from audiovisual recordings of patients' clinical interviews.
The interviews are conducted by a virtual agent to preclude human interference.
Different from the above two datasets, each modality in AVEC2019 provides several different kinds of features.
For example, acoustic modality consists of MFCC, eGeMaps, and deep features extracted by VGG~\cite{simonyan2014very} and DenseNet~\cite{huang2017densely}.
In previous researches~\cite{sun2021multi}, Hao et al. found that MFCC and AU-poses are two most discriminant features in acoustic and visual modalities, respectively.
Therefore, for the purpose of simplicity and efficiency we just employ MFCC and AU-poses features for depression detection.
The dataset is annotated by PHQ-8 scores in the interval $[0, 24]$ and bigger PHQ-8 score means the depression tendency is more severe.
There are 163 training samples, 56 validation samples, and 56 test samples in this benchmark dataset.

\subsection{Experimental Setup}
For multimodal feature extraction, we set $L$ to 100 for sentiment analysis and 1000 for depression.
Because the lengths of the samples vary, we zero pad smaller sequences and cut off longer sequences to match the length.
For each modality features, we set $D$ to 128.
In our study, $M$ is fixed to 3 because we have three involved modalities ($t$, $a$, and $v$).
For $L'$, $M'$, and $D'$, we explore them in Section~\ref{dim_explore}.
In the experiments, we find CubeMLP is a so effective structure that the performance can reach the state-of-the-art when we just set $N$ to 3.
During the training, the learning rate is initialized to 0.004 and multiplied by 0.1 every 50 epochs.
Our models are implemented with PyTorch~\cite{paszke2019pytorch} framework and validated on two GTX 1080Ti GPU cards.

\subsection{Evaluation Metrics}
\subsubsection{CMU-MOSI \& CMU-MOSEI}
The CMU-MOSI and CMU-MOSEI provide sentiment regression tasks.
Following recent works\cite{deng2021dense,chen2020swafn}, we provide MAE and Pearson correlation (Corr) as measurements.
The consecutive sentiment tendency can also be transferred to binary classification task (positive and negative) and 7-class classification task (rounded tendency, e.g., 1.8 is in class-2).
For the classification tasks, we provide accuracy (Acc) and F1-score (F1) as measurements.

\subsubsection{AVEC2019 DDS}
The CCC and MAE metrics are used to evaluate the AVEC2019 DDS dataset, which have been utilized in previous depression detection studies.
The formula expression of CCC is represented as:
\begin{equation}
  \label{ccc}
  CCC = \frac{2S_{\hat{y}y}}{S_{\hat{y}}^{2}+S_{y}^{2} + (\bar{\hat{y}} - \bar{y})^{2}}
\end{equation}
The CCC is in the interval $[-1, 1]$ and -1 represents the total negative correlation while 1 means the perfect positive correlation.

\section{Results \& Analysis} ~\label{analysis}
\subsection{Experimental Results}
Our experimental results of multimodal sentiment analysis are shown in Table~\ref{result_msa}.
Table~\ref{result_msa} also compares the results of previous studies according to their provided papers.
As is shown in the table, CubeMLP achieves a MAE of 0.770 on CMU-MOSI and 0.529 on CMU-MOSEI, which is competitive with other state-of-the-art approaches.
Among all the methods, TFN \cite{zadeh2017tensor} and MFN \cite{zadeh2018memory} are tensor-based approaches that process the sequential signals from three modalities at the same time.
ICCN \cite{sun2020learning} uses mathematical metrics to determine the relationship between modalities.
MulT \cite{tsai2019multimodal} and LMF-MulT \cite{sahay2020low} employ stacked Transformers to expand the available temporal frames for alignment.
Furthermore, MAT and MNT \cite{delbrouck2020transformer} use Transformer's self-attention mechanism to simultaneously encode two types of modality features.
SWAFN \cite{chen2020swafn} calculates the corresponding co-attention for two modality pairings (text-visual and text-acoustic) in order to share cross-modality information between modalities.
MISA \cite{hazarika2020misa} attempts to learn modality-invariant and modality-specific representations while BBFN \cite{han2021bi} learns complementary information using two symmetric Transformer-based structure.

For multimodal depression detection, our results are shown in Table~\ref{result_dds}.
On the AVEC2019 DDS dataset, we achieve a CCC of 0.583 and an MAE of 4.37.
As for other approaches, Baseline~\cite{ringeval2019avec} uses late fusion and averages the final predictions from involved modalities.
To adaptively fuse the final predictions, Sun et al. proposed the adaptive fusion Transformer network. 
Sun et al.~\cite{sun2021multi} propose the adaptive fusion Transformer network to adaptively fuse the final predictions.
EF~\cite{kaya2019predicting} introduces simple linguistic and word-duration features to estimate the depression level.
Bert-CNN \& Gated-CNN~\cite{rodrigues2019multimodal} employs the gate mechanism to fuse the information attained from involved modalities.
Multi-scale Temporal Dilated CNN~\cite{fan2019multi} uses the dilated CNN to extract multimodal features using a larger receptive field.
Hierarchical BiLSTM~\cite{yin2019multi} applies the hierarchical biLSTM to capture the sequential information in a pyramid-like structures.

\subsection{Ablation Study}
To study the effectiveness of each MLP units, we perform the ablation studies on the CMU-MOSI datasets shown in Table~\ref{tab:result_ablation}.
We can infer from the results that the sequential-mixing (MLP-L) and channel-mixing (MLP-D) play more important roles than modality-mixing (MLP-M).
The accuracy tends to get better when combining more MLPs.
The performance is best when performing the mixing at all three axis ($L$, $M$, and $D$). 

\begin{figure*}
  \centering
  \subfigure[The exploration of the selection of $L'$.]{
		\begin{minipage}[b]{0.31\textwidth}
    \includegraphics[width=1\textwidth]{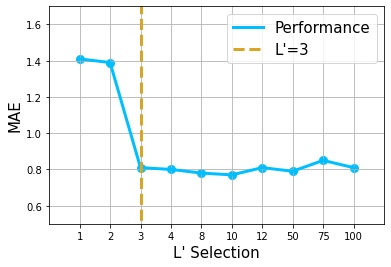}
		\end{minipage}
		\label{fig:rankl}
	}
  \subfigure[The exploration of the selection of $M'$.]{
    \begin{minipage}[b]{0.31\textwidth}
    \includegraphics[width=1\textwidth]{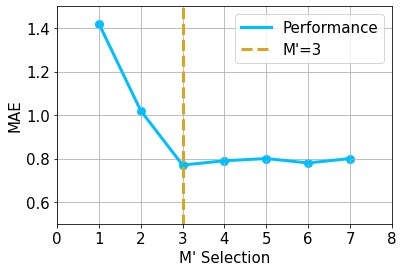}
    \end{minipage}
    \label{fig:rankm}
  }
  \subfigure[The exploration of the selection of $D'$.]{
    \begin{minipage}[b]{0.31\textwidth}
    \includegraphics[width=1\textwidth]{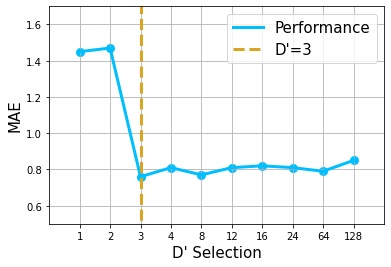}
    \end{minipage}
    \label{fig:rankd}
  }
	\caption{The output dimension ($L'$, $M'$, and $D'$) exploration of each MLP.}
	\label{fig:rank_search}
\end{figure*}

\subsection{Dimension Exploration}~\label{dim_explore}
Because of the MLPs' inherent structures, we can tune the output dimensions of each MLP ($L'$, $M'$, and $D'$).
From a mathematical point of view, $L'$, $M'$, and $D'$ represent the dimensions after the features' linear (affine) transformation on respective axis.
As a result, we investigate, how the selection affects the final prediction accuracy.
The visualization results are shown in Figure~\ref{fig:rank_search}.
It is evident that the performance decreases if the output dimensions are less than three ($<3$) on any axis. 
However, the performance fluctuation is substantially smaller if the output dimensions are $>= 3$. 
We consider that the reason for this phenomenon is the fact that we only have three modalities involved during training ($M=3$).
In other words, when treating the multimodal features as a tensor with shape $R^{L \times 3 \times D} (L >3, D>3)$, the rank of this tensor is 3.
As a result, there is little information loss while converting this tensor to $R^{3\times 3 \times 3}$ using affine transformations on all axis, resulting in constant performance but a significantly lower computing cost.

\begin{table}[]
\caption{The comparison of memory(space) consumption between CubeMLP and other state-of-the-art approaches.}
\label{result_complex}
\centering
    \begin{tabular}{c|c}
    \toprule
    Models & Space Consumption  \\
    \midrule
    TFN~\cite{zadeh2017tensor}                      & $O(L^{M})$ \\
    Adaptive Fusion Transformer~\cite{sun2021multi}       & $O(L^{2})$  \\
    ICCN\cite{sun2020learning}       & $O(L\times D^{2})$  \\
    MNT\& MAT\cite{delbrouck2020transformer}  & $O(L^{2})$ \\
    BBFN\cite{han2021bi}               & $O(L^{2})$  \\
    CubeMLP(Ours)                       & $O(max(L,M,D))$ \\
    \bottomrule
    \end{tabular}
\end{table}

\begin{figure*}[tbh]
  \centering
  \includegraphics[width=1.0\linewidth]{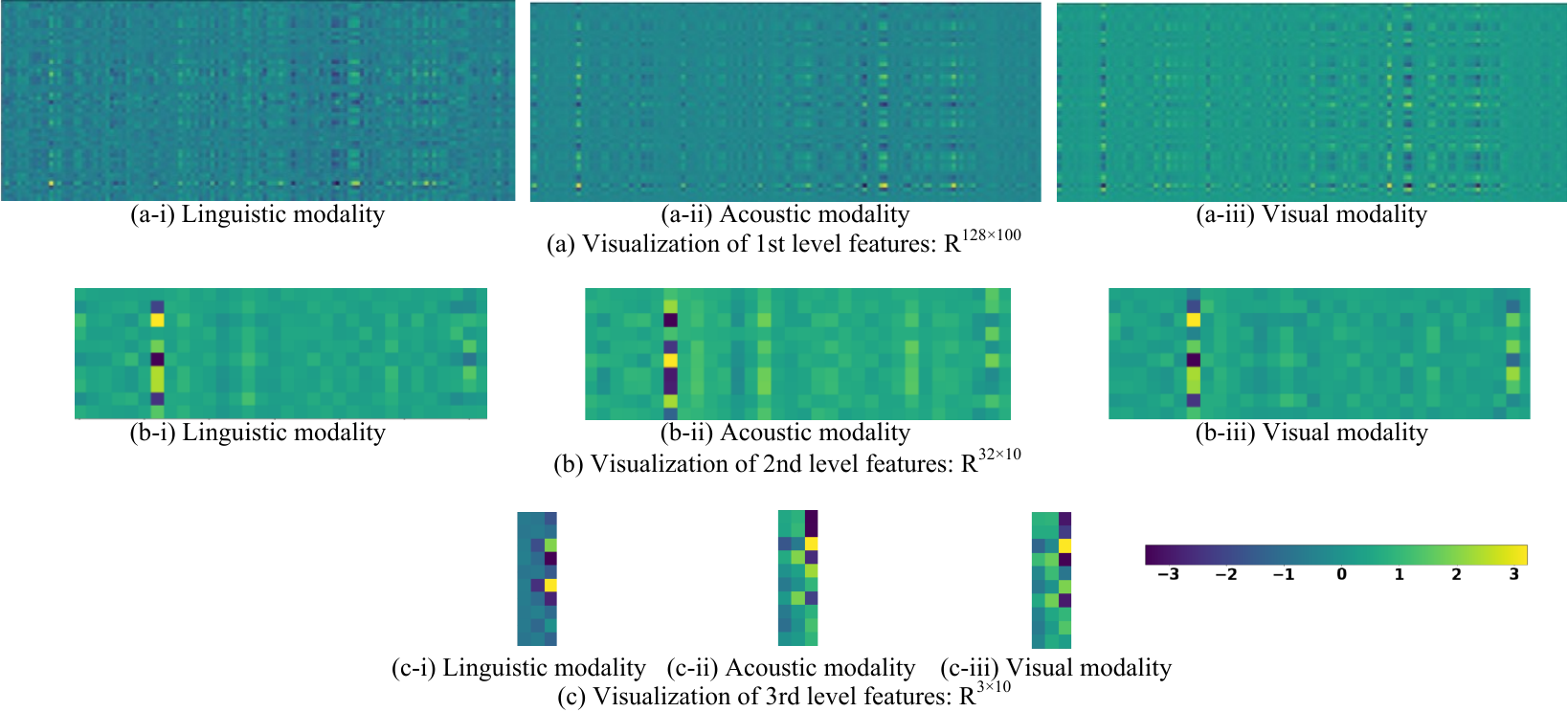}
  \caption{The visualization of each modality's features processed by CubeMLP at different levels. It can be seen that the semantic feature maps are more consistent and denser, as the model level goes deeper.}
  \Description{}
	\label{fig:features}
\end{figure*}

\subsection{Feature Visualization}
To further prove that the features of different modalities can be mapped to similar hidden spaces by CubeMLP, we visualized the features of the model on the CMU-MOSEI dataset. 
The model in this experiment is formed by three CubeMLP Blocks ($N = 3$).
For the three blocks, $D'$ is set to [128, 32, 3], while $L'$ is set to [100, 10, 10] respectively. Since CMU-MOSEI has three modalities, $M'$ is always set to 3 all along. 
The features of different CubeMLP blocks are visualized in Figure~\ref{fig:features}.
For the convenience of comparison between modalities, each $R^{L'*M'*D'}$ feature is split into $M'$ two-dimensional $L'*D'$ images.

From the visualization result, we can find that in the first CubeMLP Block, features of different modalities still have some noticeable differences between each other((a-i), (a-ii), and (a-iii)). 
However, as the feature size getting smaller and more semantic information is extracted, the output of the second CubeMLP Block tends to have similar representations for the three different modalities((b-i), (b-ii), and (b-iii)).
Finally, the third CubeMLP Block outputs condensed semantic information and the visualization of different modalities' feature become furthermore similar((c-i), (c-ii), and (c-iii)).
This indicates that after the processing of CubeMLP, the features of different modalities can be efficiently mapped to a similar hidden space.
To put it another way, the multimodal information is better transmitted and semantic features are refined, resulting in a superior modality fusion result.
With such fused information, downstream tasks can achieve a higher accuracy.

\subsection{Computational Complexity Comparison}
We compare computational space complexity of CubeMLP with other state-of-the-art techniques. The results are shown in Table~\ref{result_complex}.
One of the advantages of MLP-based structures is its low computational cost.
CubeMLP improves performance while utilizing less computational memory, demonstrating that it is a cost-efficient and effective multimodal fusion structure.

\section{Conclusion}
In this paper, we treat multimodal fusion as feature mixing and propose the MLP-based CubeMLP for unified multimodal feature processing.
In CubeMLP, we perform the mix-up at all axis of multimodal features.
CubeMLP can reach the state-of-the-art performance for sentiment analysis and depression detection while keeping the computational burden low.
We analyzed CubeMLP's components and compared it to other techniques.
Not only that, We conducted extensive ablation studies and visual analysis to illustrate MLP's efficiency and multimodal processing capabilities.
We intend to do more MLP research in the future, as well as more experiments in other multimodal fusion domains, such as emotion detection.
MLP, we believe, will become a widely used solution for a variety of tasks.

\begin{acks}
Thanks to Rahul for the careful English proofing and language grammar check.
Thanks to the paper reviewers for the constructive research comments.
\end{acks}

\bibliographystyle{ACM-Reference-Format}
\balance
\bibliography{sample-base}


\end{document}